\documentclass[12pt]{iopart}
\usepackage{graphicx}
\usepackage{latexsym}
\usepackage{amsfonts,amsthm}

\begin{document}
\title{Low-density series expansions for directed percolation IV.  
Temporal disorder}
\author{Iwan Jensen}
\address{
ARC Centre of Excellence for Mathematics and Statistics of Complex Systems, 
Department of Mathematics and Statistics, 
The University of Melbourne, Victoria 3010, Australia}

\date{\today}

\ead{I.Jensen@ms.unimelb.edu.au}

\submitto{\JPA}


\begin{abstract}
We introduce a model for temporally disordered directed percolation in
which the probability of spreading from a vertex $(t,x)$, where $t$ is the
time and $x$ is the spatial coordinate, is independent of $x$ 
but depends on $t$. Using a very efficient algorithm we calculate low-density 
series for bond percolation on the directed square lattice. Analysis of the series 
yields estimates for the critical point $p_c$ and various critical exponents which 
are consistent with a continuous change of the critical parameters as the strength 
of the disorder is increased. 
\end{abstract}

\maketitle

\section{Introduction \label{sec:intro}}

Directed percolation (DP) can be thought of simply as a percolation process 
on a directed lattice in which connections are allowed only in a preferred 
direction, e.g., on hyper-cubic lattices connections are only allowed along 
edges connecting vertices with increasing coordinates. DP is most commonly 
interpreted as a growth model and the preferred direction $t$ is time. 
Bond percolation is also a special case of a $(d+1)$-dimensional stochastic cellular 
automaton \cite{DK84a}. On the square lattice the evolution  
is governed by the transition probabilities $W(\sigma_x | \sigma_l,\sigma_r)$,
with $\sigma_i = 1$ if the vertex $i$ is occupied and 0 otherwise, which is the 
probability of finding the vertex $x$ in state $\sigma_x$ at time $t$ given that, 
at time $t-1$, the vertices $x-1$ and $x+1$ were in states $\sigma_l$ and
$\sigma_r$, respectively. Bond percolation corresponds to the choice
$W(0| \sigma_l,\sigma_r) = (1-p)^{\sigma_l+\sigma_r}$.
The behavior of the model is controlled by the spreading
probability (or density of bonds) $p$. When $p$ is smaller 
than a critical value $p_c$ all clusters remain finite, and in this 
sub-critical region any initial population will eventually die out. Above $p_c$
there is a non-zero probability of finding an infinite cluster,
the average cluster-size $S(p)$ diverges as $p \rightarrow p_c$,
and the initial population will increase exponentially fast. 

Directed percolation, and a closely related continuous time version the 
contact process \cite{TEHarris74},
serve as the prime examples of models for population growth exhibiting a 
non-equilibrium phase transition into an absorbing state (a state from which the 
system cannot escape), in these cases the state totally devoid of occupied 
sites. DP type transitions are also encountered in many other situations,
perhaps most prominently in models for chemical reactions \cite{Schlogl72,Grassberger79a} 
including heterogeneous catalysis and surface reactions \cite{ZGB86}.
For a recent comprehensive review see \cite{Hinrichsen00a}

Most studies of non-equilibrium systems have been limited to cases
without disorder, that is the spreading probability is homogeneous
in both time and space. Obviously in many real systems this idealisation
is unrealistic. Often some degree of disorder is present, e.g., in a
real catalyst some sites may be blocked by impurities leading one to
consider models with quenched spatial disorder. Likewise in trying
to model real growth processes one would like to consider
temporal disorder so as to model changes in conditions from
year to year, seasonal changes etc. 

Non-equilibrium models with quenched spatial disorder were considered
by Kinzel \cite{Kinzel85a}. He argued that the Harris criterion \cite{ABHarris74},
well-known from disordered spin-systems, should apply for DP as well. 
Harris argued that in systems with quenched disorder one must have
$d\nu \geq 2$, where $\nu$ is the correlation length exponent.
This means that quenched disorder is a relevant perturbation and thus
should change the critical exponents if in the pure system $d\nu<2$.
The Harris criterion has been rigorously established for a large class of 
disordered systems by Chayes \etal \cite{CCFS86}.
In the case of DP the Harris criterion becomes $d\nu_{\perp} < 2$,
and quenched disorder is a relevant perturbation for $d \leq 3$. The effect of 
quenched disorder was studied numerically by Noest \cite{Noest86,Noest88} who
found a marked changed in the static critical exponents independent of the strength
of the disorder. 
Moreira and Dickman \cite{Dickman96a,Dickman98a} used Monte Carlo simulations to study 
the site diluted two-dimensional contact process on the square lattice.
They found a marked change in the static critical exponents $\beta$ and $\nu_{\perp}$
with a value of $\nu_{\perp}=1.00(9)$ consistent with the Harris criterion.
The dynamic behaviour was found to be incompatible with the usual scaling
observed in pure models. In particular critical spreading was found to be logarithmic
rather than power law, and the survival probability in the sub-critical regime
decayed algebraically rather that exponentially. This means that the dynamical critical
correlation length exponent $\nu_{||}$ is undefined. 
Janssen \cite{Janssen97a} investigated the problem using renormalisation group
theory and confirmed the findings of Moreira and Dickman. 
Recently, Hooyberghs, Igl\'oi and Vanderzande \cite{HIV03,HIV04} have investigated
these problems using a real-space renormalisation group framework and the 
numerical technique of the density matrix renormalisation group. They found
that for strong enough disorder the behaviour is controlled by a strong disorder
fixed point, with logarithmic dynamical correlations and  critical exponents
different from the pure system. For weaker disorder the numerical evidence 
was consistent with continuously varying static exponents.

Kinzel \cite{Kinzel85a} also briefly considered the effect of temporal disorder
and argued that a Harris type criterion applied but with $d\nu_{\perp}$ replaced
with $\nu_{||}$. Since  $\nu_{||}=1.733847(6)$ \cite{IJ99a} for (1+1)-dimensional
DP this would make temporal disorder a relevant perturbation. In a previous paper 
\cite{IJ96b} we studied temporally disordered directed percolation on the square 
lattice using a simple model in which spreading from some rows was {\em deterministic} 
(that is spreading takes place with probability 1) while spreading from the other rows
took place with probability $p$ as in the pure model. Disorder was introduced
by letting any given row be ``deterministic'' with probability $\alpha$ independent
of other rows. The model was studied using high-density series expansions for the 
percolation probability and Monte Carlo simulations of growing clusters from a single 
seed. The major finding was that the critical exponents changed continuously with the 
strength of the disorder. In particular we found that $\nu_{||}<2$ over a wide range of
values in apparent violation of the Harris criterion. 

The model used in \cite{IJ96b}
is not suitable for a low-density expansion so in this paper we study another 
model of DP with temporal disorder. The spreading probability $p(t)$ is chosen 
at random to be either $p$ or $\alpha p$ with probability $\frac12$. 
By varying the parameter $\alpha$ we can vary the strength of the disorder.
There is an obvious symmetry between the two regions $\alpha < 1$ and
$\alpha>1$ and we shall therefore only consider the case $\alpha>1$. 
Analysis of the series indicate that for any given value of $\alpha$ 
there is critical point $p_c(\alpha)$ where the system has a phase transition.
For values of the spreading probability $p$ close to $p_c(\alpha)$ the model 
changes from periods of sub-critical to periods of super-critical growth.
For large values of $\alpha$ there is a very pronounced difference between the
two regimes and the population dynamics changes from feast to famine. In the
infinite lattice limit the system will have arbitrarily long stretches of
`famine' conditions in which the population is rapidly reduced and very
likely do die out. Note that we can be sure that the model will have
sub- and super-critical regions. We can simply choose $p>p_c$, where
$p_c$ is the critical point of the pure model, and we will always have 
a super-critical growth process,
while for $\alpha p< p_c$ the growth process is always sub-critical. 

In Section~\ref{sec:calc} we give a description of the method
used to derive the low-density series for this model. The series is then 
analysed (for various values of $\alpha$) and the results presented
in Section~\ref{sec:ana}. Finally Section~\ref{sec:sum} contains a brief
summary and discussion of the main results.

\section{Calculation of low-density series \label{sec:calc}}

In the low-density phase ($p<p_c$) many quantities of interest can be 
derived from the pair-connectedness $C_{t,x}(p)$, which is the probability 
that the vertex at position $x$ is occupied at time $t$ given that the
origin was occupied at $t=0$.  
Of particular interest are moments of the pair-connectedness

\begin{equation}
  \label{eq:momdef}
  \mu_{m,n}(p) = \sum_t \sum_x t^m x^n C_{t,x}(p)
\end{equation}
\noindent
Due to symmetry moments involving odd powers of $x$ are identically zero.
The remaining moments diverge as $p$ approaches $p_c$ from below

\begin{equation}
  \label{eq:momdiv}
  \mu_{m,n}(p) \propto (p_c-p)^{-(\gamma+m\nu_{\parallel}+n\nu_{\perp})},
  \;\;\;\; p \rightarrow p_c^-
\end{equation}
\noindent

In a previous paper \cite{IJ99a} we gave a detailed description of how the 
graph theoretical properties of the pair-connectedness \cite{AE77} can
be turned into a very efficient algorithm for the calculation of low-density
series expansions for ordinary directed percolation. The series expansions for 
the disordered model is a simple generalisation of this work. Before describing
the algorithm for the disorder system we will briefly review the pure case.

It has been shown \cite{AE77} that the pair-connectedness can be expressed as a 
sum over all graphs (or finite clusters) formed by taking unions
of directed paths connecting the origin to the vertex $(x,t)$, 
\begin{equation}
  \label{eq:pairconn}
  C_{t,x}(p)= \sum_g d(g)p^{|g|},
\end{equation}
\noindent
where $|g|$ is the number of bonds in $g$. The weight $d(g)=(-1)^{c(g)}$,
where $c(g)$ is the cyclomatic number of the graph $g$. Note that $c(g)$
is increased by 1 whenever two paths join, e.g., if there are two incoming
bonds on a vertex. The restriction to unions of paths is very strong and  one 
immediate consequence is that graphs with dangling parts make no contribution 
to $C_{t,x}$ and any contributing graph terminates exactly at $(t,x)$. Another 
way of stating the restriction is that any vertex with an incoming bond {\em must} 
have an outgoing bond unless it is the terminal vertex $(t,x)$.
Any directed path to a vertex whose parallel distance 
from the origin is $t$ contains at least $t$ bonds. 
One can do much better by  using a so-called non-nodal graph
expansion \cite{EGDB88a} to extend the series to order $2t$. 
A graph $g$ is nodal if it has a vertex (other than the terminal
vertex) through which all paths pass. It is clear that each such nodal
point effectively works as a new origin for the cluster growth, and
we can obviously obtain any contributing graph by concatenating non-nodal
graphs. Note that the graph consisting of a single bond is non-nodal so all
linear graphs can be obtained by repeated concatenations. This is
the essential idea behind the non-nodal graph expansion, which proceeds
in two principal steps. First we calculate the contribution $C^*_{t,x}$
of non-nodal graphs to the pair-connectedness. The calculation is done
up to a preset order $N$, where $N$ is limited by the available computational
resources (primarily physical memory). Next we use repeated
concatenation operations of $C^*_{t,x}$ to calculate the pair-connectedness
$C_{t,x}$ and from this we finally calculate various moments $\mu_{m,n}(p)$.

\begin{figure}
\begin{center}
\includegraphics[scale=0.9]{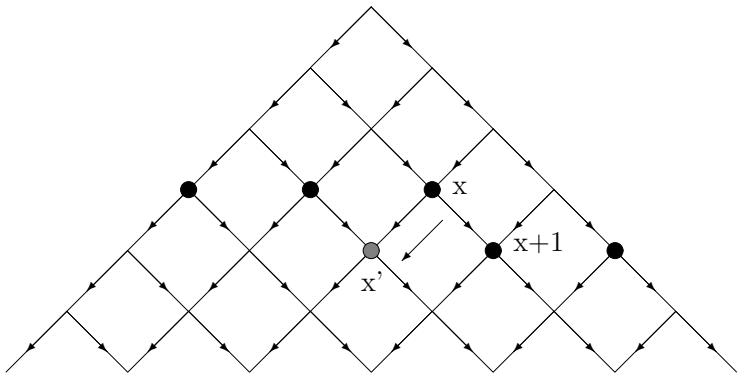}
\end{center}
\caption{\label{fig:transfer}
Example of how the boundary line (cutting through the filled circles) is
moved in order to insert another vertex (shaded circle).} 
\end{figure}

The calculation of $C^*_{t,x}$ is done efficiently using transfer-matrix
techniques. This involves drawing a boundary line across a finite
slice of the lattice and then moving the boundary line such that
one adds row after row with each row built up one vertex at a time,
as illustrated in figure~\ref{fig:transfer}.
The sum over all contributing graphs is calculated as the lattice is
constructed. At any given stage the boundary line cuts through a
number of, say $j$, vertices. There are two possible states (0 or 1) per 
vertex, corresponding to vertices with (1) or without (0) incoming
bonds, leading to a total of $2^j$ possible boundary configurations. 
However, as explained in \cite{IJ99a} not all the $2^j$ possible configurations
are required. Many of them can be discarded because they either don't
make a contribution to  $C^*_{t,x}$ or only contributes above order $N$.
The weight of each configuration is given by a polynomial $P$ in $p$
truncated at order $N$. The updating of $P$, as the boundary line
is moved to insert a new vertex, depends only on the states of the vertices
marked $x$ and $x+1$ in figure~\ref{fig:transfer} and are given simply by

\begin{eqnarray} \label{eq:update}
P({\overline{S}}_{1,1} )& = & p^2 P(S_{1,0}) + pP(S_{1,1}) - p^2P(S_{1,1}), \nonumber \\
P({\overline{S}}_{0,1}) & = & p P(S_{1,0}) + P(S_{0,1}) - pP(S_{1,1}), \nonumber \\
P({\overline{S}}_{1,0}) & = & p P(S_{1,0}), \\
P({\overline{S}}_{0,0}) & = &   P(S_{0,0}). \nonumber 
\end{eqnarray}
Here $S_{a,b}$ is the configuration before the move which has the vertex at $x$ 
in state $a$ and the vertex at $x+1$ in state $b$, while $\overline{S}_{a,b}$ is 
the configuration after the move  which has the new vertex at $x'$ in state 
$a$ and the vertex at $x+1$ in state $b$. For a derivation of the updating rules 
see \cite{IJ99a}.

Limiting the calculation to non-nodal contributions is very simple; 
whenever the boundary line reaches the horizontal position one just
sets to zero the polynomials of states with a single incoming bond.
This obviously ensures that configurations with a nodal point
are deleted from the calculation. The pair-connectedness at the
following time can be calculated from the states with incoming
bonds at nearest neighbour sites and no incoming bonds on any other sites. 
In this way we calculate the two non-nodal series 
$C_t^* (p) = \sum_{x} C_{t,x}^* (p)$ and $X^*_t(p) = \sum_{x} x^2 C_{t,x}^* (p)$. 

The generalisation to DP with temporal disorder is quite simple. First of
all the configuration weights $P$ will depend on two variables $u$ and $v$
corresponding to the two spreading probabilities. Lets consider the situation
in which a row has just been completed. Since the probability is $\frac12$ 
that spreading from a given row occurs with probability $u$ or $v$, respectively,
obviously $P(u,v)=P(v,u)$. Next we insert a new row. Formally we can express
the weight of any new configuration $\overline{S}$ as a weighted sum over the 
weights of the configurations $S$ in the row above
\begin{equation}
P_{\overline{S}}(u,v) = \sum_S W(u)P_S(u,v) + \sum_S W(v)P_S(u,v).
\end{equation}
From the nature of the problem it is clear that the weight $W$ is the
same (apart from the change of variable). This means that it is very
simple to calculate the non-nodal expansion. Having completed a row,
we just use the updating rules in equation~(\ref{eq:update}) with the
spreading fixed at $u$. This gives of a set of incomplete weights
\begin{equation}
Q_{\overline{S}}(u,v) = \sum_S W(u)P_S(u,v).
\end{equation}
But since $P_S(u,v)$ is symmetric in $u$ and $v$ we get the desired
final result as
\begin{equation}
P_{\overline{S}}(u,v) = Q_{\overline{S}}(u,v)+Q_{\overline{S}}(v,u).
\end{equation}
Note that this is a very efficient algorithm. Naively one would have
expected that to calculate the pair-connectedness one would have to
sum over all $2^t$ realisations of the disorder. However, as
we have just seen this is not necessary, the only complication being
the requirement to maintain a two parameter generating function.
Ultimately we shall calculate series  for $\mu_{m,n}(p)$, by
fixing a value of $\alpha$ and setting $u=p$ and $v=\alpha p$.
So to need only retain the coefficients of $u^iv^j$ provided
$i+j \leq N$. The first few
terms in the expansion for $C_t^*(u,v)$ are:
\begin{eqnarray*}
\fl C_1^*(u,v) = 2(u+v) \\
\fl C_2^*(u,v) = -u^4-2u^2v^2-v^4 \\
\fl C_3^*(u,v) = 2u^7-2u^6+(4u^5-6u^4)v^2+2u^4v^3+(2u^3-6u^2)v^4+4u^2v^5-2v^6+2v^7 \\ 
\fl C_4^*(u,v) = u^{12}-4u^{11}+8u^9-5u^8 + (2u^{10}-8u^9+24u^7-20u^6)v^2 -(4u^8-8u^6)v^3 \\
\lo{+(3u}^8-8u^7+24u^5-30u^4)v^4 -(8u^6-24u^4)v^5+(4u^6-8u^5+8u^3-20u^2)v^6 \\
\lo{-(8u}^4-24u^2)v^7+(4u^3-4u-5)v^8-(8u^2-8)v^9+2u^2v^{10}-4v^{11}+v^{12}
\end{eqnarray*}
Note that $C_t^*(p,p)$ is $2^tC_t^*(p)$, where $C_t^*(p)$ is the non-nodal
pair-connectedness of the pure model. In order 
to calculate our final series for the average cluster size $S(p)$
and the moments $\mu_{1,0}(p)$ and $\mu_{2,0}(p)$ we
simply fix a value of $\alpha$. We then use repeated concatenations
of $C_t^*(p,\alpha p)/2^t$ to calculate the pair-connectedness 
$C_t (p)$ (with $\alpha$ fixed), which in turn we use to calculate
the moments. The series for the second transverse moment $\mu_{0,2}$
is obtained as  $\mu_{0,2} = S^2(p)\sum_t X^*_t(p)$.

The series for $C_t^*(u,v)$ was derived correctly to order 115 (that is all
coefficients of $u^iv^j$ were calculated exactly for $i+j\leq 115$). This
obviously results is series for $S(p)$ and the other moments to order 115 as
well. The algorithm used up to 5Gb of physical memory and the total CPU time
was about 30 days.

\section{Analysis of series \label{sec:ana}}

The various series were analysed using inhomogeneous differential 
approximants \cite{AJG89a}.  Suffice to say that a $K$th-order
differential approximant to a function $f$ is a solution to an 
inhomogeneous differential equation
\begin{equation}\label{eq:diffapp}
\sum_{i=0}^K Q_{i}(x)(x\frac{\mbox{d}}{\mbox{d}x})^i \tilde{f}(x) = P(x),
\end{equation}
where the coefficients in the polynomials $Q_i$ and $P$ of order $N_i$ and $L$, 
respectively, are chosen so that the series for the function $\tilde{f}(x)$, 
agrees with the series coefficients of $f$. The equations are
readily solved as long as the total number of unknown coefficients in
the polynomials is smaller than the order of the series $n$.
The possible singularities of the series appear as
the zeros $x_i$ of the polynomial $Q_K$ and the associated critical
exponent $\lambda_i$ is estimated from the indicial equation.

\begin{figure}
\begin{center}
\includegraphics[scale=0.65]{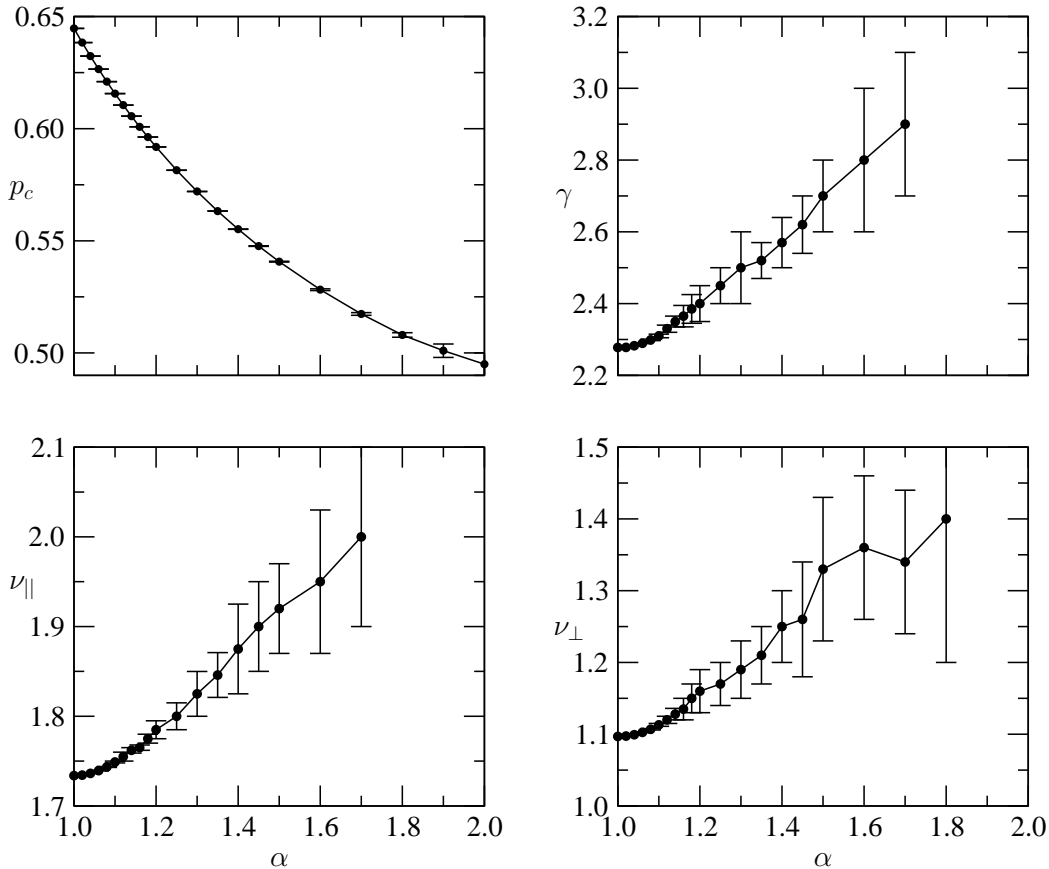}
\end{center}
\caption{\label{fig:crpexp}
Estimates for the critical point $p_c$ and critical exponents $\gamma$,
$\nu_{||}$ and $\nu_{\perp}$ as a function of the disorder strength $\alpha$.} 
\end{figure}

\begin{table}
\begin{indented}
\item[]\caption{\label{tab:crpexp} 
Estimates for the critical point $p_c$ and critical exponents $\gamma$,
$\nu_{||}$ and $\nu_{\perp}$ at various values of the disorder strength $\alpha$.}
\begin{tabular}{lllll}
\br
$\alpha$ & $p_c$ & $\gamma$ & $\nu_{||}$ & $\nu_{\perp}$ \\ 
\mr
1.00  &  0.6447002(2)  &  2.27765(5)  &  1.73385(5)  &  1.09687(1) \\
1.02  &  0.6383905(5)  &  2.2779(1)   &  1.73442(8)  &  1.09745(5) \\
1.04  &  0.6323450(10) &  2.2825(5)   &  1.73635(15) &  1.0993(2)  \\
1.06  &  0.6265485(15) &  2.2895(10)  &  1.7394(3)   &  1.1025(5)  \\
1.08  &  0.6209865(25) &  2.299(2)    &  1.7435(5)   &  1.1068(10) \\
1.10  &  0.615645(4)   &  2.310(5)    &  1.748(2)    &  1.1135(25)  \\
1.20  &  0.59183(3)    &  2.40(5)     &  1.785(10)   &  1.16(3)    \\
1.30  &  0.57200(6)    &  2.5(1)      &  1.825(25)   &  1.19(4)    \\
1.40  &  0.55520(10)   &  2.57(7)     &  1.875(50)   &  1.25(5)    \\
1.50  &  0.5407(2)     &  2.7(1)      &  1.92(5)     &  1.33(10)   \\
1.60  &  0.5282(4)     &  2.8(2)      &  1.95(8)     &  1.36(10)   \\
1.70  &  0.5174(6)     &  2.9(2)      &  2.0(1)      &  1.34(10)   \\
\br
\end{tabular}
\end{indented}
\end{table}

Our use of differential approximants for series analysis has been detailed in 
previous papers \cite{IJ99a,IJ96a} and the interested reader can refer
to these papers and the comprehensive review \cite{AJG89a} for further details.
Typically we obtain estimates for $p_c$ and the critical exponents by averaging 
values obtained from second and third order differential approximants. 
For each order $L$ of the inhomogeneous polynomial we average over those 
approximants which use at least the first $90\%$ of the terms in the series.
A rough error estimate is obtained from the spread among the approximants. Note 
that these error bounds should {\em not} be viewed as a measure of the true error 
as they cannot include possible systematic sources of error.

In figure~\ref{fig:crpexp} we have plotted the estimates for the critical 
point $p_c$ and the critical exponents $\gamma$, $\nu_{||}$ and $\nu_{\perp}$ 
as a function of the disorder strength $\alpha$ (some of the estimates are 
listed explicitly in Table~\ref{tab:crpexp}). The estimates were obtained 
by analysing the series $S(p)\propto (p-p_c)^{-\gamma}$, 
$\mu_{2,0}(p)/\mu_{1,0}(p)\propto (p-p_c)^{-\nu_{||}}$ and 
$\mu_{0,2}(p)/S(p)\propto (p-p_c)^{-2\nu_{\perp}}$, respectively, and 
are based on results using both second and third order differential approximants
with varying degrees of the inhomogeneous polynomial. The results are clearly
compatible with a continuous change in all the critical parameters as the strength of 
the disorder $\alpha$ increases. Naturally we observe
a decrease in the value of the critical point $p_c$, and in particular we
note that $\alpha p_c \leq 1$. The critical exponents all increase with
increasing $\alpha$. In particular we note that $\nu_{||} < 2$, at least
while $\alpha \leq 1.5$. It is also clear that the change in the values of the 
critical exponents is larger than the estimated error bars, clearly indicating that the
change is not due merely to less well behaved series. When the disorder is relatively 
weak ($\alpha$ close to 1) the estimates for both $p_c$ and the exponents are still
quite sharp, but as $\alpha$ is increased the estimated error bars become large 
and for values of $\alpha \geq 1.5$ the exponent estimates are inaccurate.

\begin{table}
\caption{\label{tab:ce110} 
Estimates of the critical point $p_c$ and critical exponents $\gamma$,
$\nu_{||}$ and $2\nu_{\perp}$ with $\alpha=1.1$, as obtained
from third order differential approximants ($L$ is the order of the
inhomogeneous term).}
\begin{center}
\footnotesize
\begin{tabular}{lllllll}
\br
$L$ & $p_c$ & $\gamma$ & $p_c$ & $\nu_{\parallel}$ & $p_c$ & $2\nu_{\perp}$ \\ 
\mr
0  & 0.6156457(10) & 2.3114(29) & 0.615644(71)  & 1.83(11)   & 0.6156455(42) & 2.2275(43) \\
5  & 0.6156454(17) & 2.3115(17) & 0.615634(35)  & 1.765(23)  & 0.6156474(25) & 2.2294(29) \\
10 & 0.6156399(16) & 2.3070(15) & 0.6156464(23) & 1.7489(11) & 0.6156449(28) & 2.2268(29) \\
15 & 0.6156437(22) & 2.3101(18) & 0.6156427(82) & 1.7491(30) & 0.6156458(28) & 2.2277(30) \\
20 & 0.6156460(26) & 2.3121(25) & 0.6156407(85) & 1.7483(26) & 0.6156447(79) & 2.2270(75) \\
\br
\end{tabular}
\end{center}
\end{table}

\begin{figure}
\begin{center}
\includegraphics[scale=0.65]{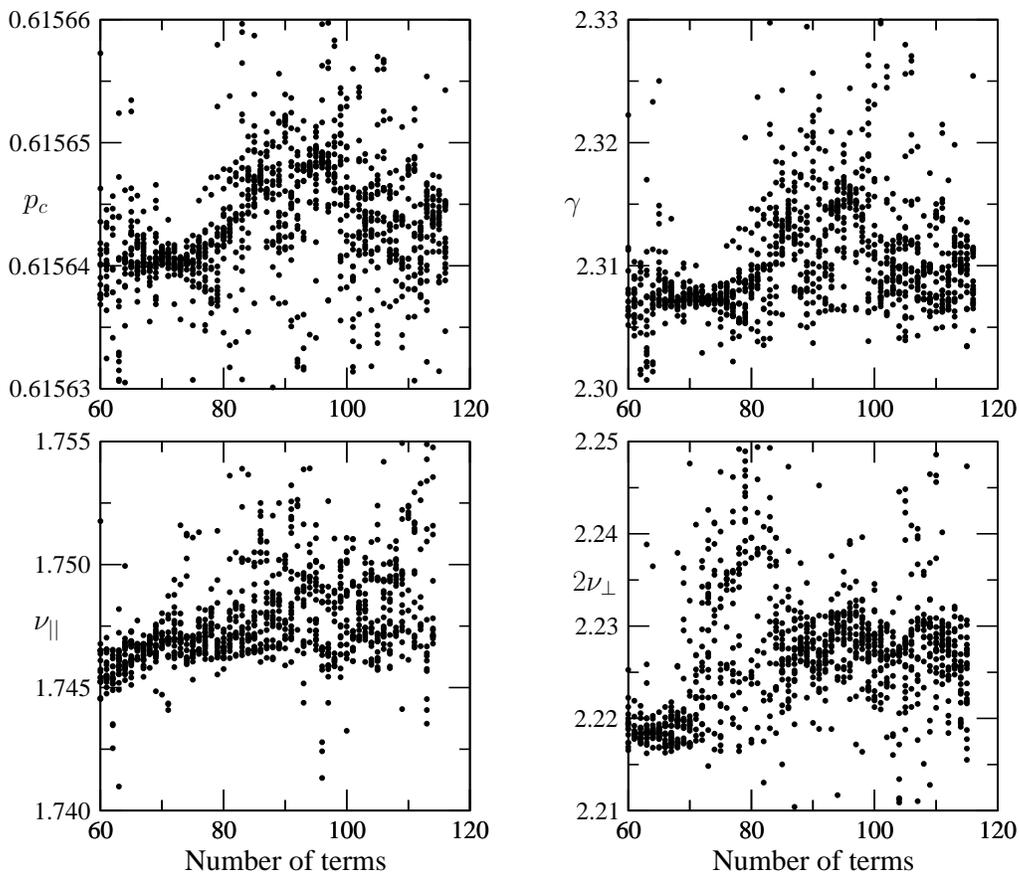}
\end{center}
\caption{\label{fig:ce110}
Estimates of the critical point $p_c$ and critical exponents $\gamma$, 
$\nu_{||}$ and $2\nu_{\perp}$ with $\alpha=1.1$ plotted versus the number of 
terms used by the third order differential approximants.} 
\end{figure}

As a concrete example of the analysis we show some detailed results from the
analysis with the particular value $\alpha=1.1$. In Table~\ref{tab:ce110} 
we have listed estimates for $p_c$ and the critical exponents $\gamma$, 
$\nu_{||}$ and $2\nu_{\perp}$. The estimates of $p_c$ are consistent
to 5 digits both among the three series as well as among the approximants
using different orders of the inhomogeneous polynomial. Likewise the
estimates for the exponents are consistent as we vary the order of
the inhomogeneous polynomial (the only exception being the $L=0$ and $L=5$
estimates of $\nu_{\parallel}$). Taking into account the spread among
the different sets of approximants (while ignoring the obviously spurious
results for  $\nu_{\parallel}$) we arrive at the estimates listed in 
Table~\ref{tab:crpexp}.
To gauge whether there are pronounced sources of systematic errors we plot 
in figure~\ref{fig:ce110} the estimates for the critical point $p_c$ and the 
critical exponents $\gamma$, $\nu_{||}$ and $2\nu_{\perp}$ as a function of the 
number of terms used by the differential approximants. The critical point 
estimates are those obtained from the cluster size series $S(p)$. Each point 
in these plots corresponds to the estimate obtained from a single third order 
differential approximant. From these plots we observe that the estimates
do not change much as the number of terms is increased and thus conclude that 
there does not appear to be any systematic errors in the estimates.

\section{Summary \label{sec:sum}} 

We have introduced a simple model for directed percolation with temporal
disorder, described a very efficient algorithm for the derivation of
low-density series expansions for this model and presented results
from the analysis of the series. The main result from the series analysis
is that the model has a critical point $p_c (\alpha)$ which varies
continuously with $\alpha$ as does the critical exponents $\gamma$,
$\nu_{\parallel}$ and $\nu_{\perp}$. The estimates for the
critical exponent $\nu_{\parallel} < 2$ for most values of the disorder
in apparent violation of the Harris criterion.

\section*{E-mail or WWW retrieval of series}

The series for the directed percolation problem studied in this paper 
can be obtained via e-mail by sending a request to 
I.Jensen@ms.unimelb.edu.au or via the world wide web on the URL
http://www.ms.unimelb.edu.au/\~{ }iwan/ by following the relevant links.

\section*{Acknowledgments}

The calculations presented in this paper were in part performed on
the facilities of the Australian Partnership for Advanced Computing (APAC) 
and the Victorian Partnership for Advanced Computing (VPAC). 
We gratefully acknowledge financial support from the Australian Research Council.

\section*{References}


\end{document}